\documentclass[pra, twocolumn, amsmath, amssymb, showpacs, unsortedaddress, floatfix]{revtex4-1}

\usepackage{graphicx, color}

\newcommand{\ud}{\ensuremath{\:\mathrm{d}}}

\newcommand{\ket}[1]{\ensuremath{\vert #1 \rangle}}
\newcommand{\bra}[1]{\ensuremath{\langle #1 \vert}}
\newcommand{\bracket}[2]{\ensuremath{\langle #1 \vert #2 \rangle}}
\newcommand{\ketbra}[2]{\ensuremath{\vert #1 \rangle\langle #2 \vert}}

\let\vec\mathbf

\begin{document}

\title{Towards an experimentally feasible controlled-phase gate on two
  blockaded Rydberg atoms}

\author{Matthias M. M\"{u}ller}
\email{matthias.m.mueller@uni-ulm.de}
\author{Michael~Murphy}
\author{Simone~Montangero}
\author{Tommaso~Calarco}
\affiliation{Institut f\"{u}r Quanteninformationsverarbeitung, Universit\"{a}t Ulm, Albert-Einstein-Allee 11, D-89069 Ulm, Germany}
\author{Philippe~Grangier}
\author{Antoine~Browaeys}
\affiliation{Laboratoire Charles Fabry, Institut d'Optique, CNRS, Univ Paris-Sud, Campus Polytechnique, RD 128, 91127 Palaiseau cedex, France}

\begin{abstract}
  We investigate the implementation of a controlled-Z gate on a pair of
  Rydberg atoms in spatially separated dipole traps where the joint
  excitation of both atoms into the Rydberg level is strongly suppressed
  (the Rydberg blockade). We follow the adiabatic gate scheme of
  Jaksch~\emph{et al.} \cite{Jaksch2000}, where the pair of atoms are
  coherently excited using lasers, and apply it to the experimental
  setup outlined in Ga\"{e}tan~\emph{et al.} \cite{Gaetan2009}. We apply
  optimisation to the experimental parameters to improve gate fidelity,
  and consider the impact of several experimental constraints on the
  gate success.
\end{abstract}


\maketitle


\section{Introduction}
Using neutral atoms for quantum information has garnered much
theoretical interest over the last decade, fuelled by advances in their
experimental manipulation, particularly trapping and cooling. Several
novel schemes for entangling pairs of atoms (an essential operation for
quantum logic) via controlled collisions have been developed
\cite{Jaksch1999, Calarco2000}, but schemes that make use of the special
properties of Rydberg atoms are also very promising (see
\cite{Saffman2010} for a review). In particular, several schemes for
producing quantum gates by exciting pairs of Rydberg atoms with tuned
lasers have emerged \cite{Jaksch2000, Moller2008} which capitalise on
the strong dipole-dipole interaction that prevents the simultaneous
excitation of neighbouring Rydberg atoms, known as the Rydberg
blockade. Several steps towards realising such schemes experimentally
have already been achieved, particularly the observation of the blockade
\cite{Gaetan2009} and entanglement generation \cite{Wilk2010} in a
system of two confined Rydberg atoms. There has even been some early
success in producing a gate with trapped Rydberg atoms
\cite{Isenhower2010}.

In this paper, we consider the implementation of a controlled-Z
(\textsc{cz}) gate on a pair of Rydberg atoms confined in spatially
separated dipole traps subject to the Rydberg blockade effect. We follow
the scheme outlined in \cite{Jaksch2000}, but with specific application
to the experimental setup detailed in \cite{Miroshnychenko2010}, where
the Rydberg atom is excited via a two-photon transition. This proposal
has the advantage that both atoms are excited by the same laser,
reducing the need for single-atom addressability; the gate is also
adiabatic, which softens the experimental requirement for strong fields
or precise timings. However, the experimental considerations do present
additional challenges in the implementation of the gate, particularly
due to loss from the intermediate state of the transition and the
movement of the atoms in the dipole trap. We will attempt to address
both of these issues here by applying a direct search
control algorithm to search for the ideal set of parameters for implementing the
gate on a short timescale ($\sim 1\ \mu$s) and with high fidelity. Our
results will show that the physical system allows for a great deal of
control and gate times and fidelities approaching our desired range,
providing a positive outlook for implementing high-fidelity gates with
such systems.

The paper is arranged as follows. In Sec.~\ref{sec:cphase}, we briefly
recount the \textsc{cz} gate, followed by a description of how it may be
synthesised on a pair of Rydberg atoms. In particular, we expose the
operation of the gate by considering the effective two-level dynamics of
each atom and its interaction with a laser field. In Sec.~\ref{sec:opt},
we describe how we optimise the operation of the laser using a gradient
descent to achieve the gate with high-fidelity. In Sec.~\ref{sec:exp},
we consider the details of the experiment and the constraints it imposes
on the gate operation, particularly with regards to loss and effects
arising from atomic motion. Finally, we conclude our paper in
Sec.~\ref{sec:conc}.


\section{Controlled phase gate}\label{sec:cphase}

\subsection{Gate definition}
The controlled-Z (\textsc{cz}) gate is a two-qubit gate in quantum
information, and belongs to the class of controlled unitary
operations \cite{Barenco1995}. Given the computational basis
${\ket{0},\ket{1}}$, it is defined as the unitary transformation
\begin{equation}\label{eqn:cnotmat}
  \mathrm{CZ} = \begin{pmatrix}
    1 & 0 & 0 & 0 \\
    0 & 1 & 0 & 0 \\
    0 & 0 & 1 & 0 \\
    0 & 0 & 0 & -1
  \end{pmatrix}\,.
\end{equation}
This gate is of particular importance because it can generate entanglement
between two unentangled qubits depending on the initial states of the qubits. In addition, together with a finite set
of single-qubit operations, one can construct any desired gate operation
simply by taking combinations of these operations with the \textsc{cz}
gate. This is known as a universal set for quantum computation
\cite{Barenco1995a}.

\subsection{Blockaded Rydberg atoms} 

The physical system we are considering for the implementation of the gate is a
pair of trapped Rydberg atoms \cite{Gaetan2009}, specifically ${}^{87}$Rb. The
atoms are trapped a distance $r$ apart in two separate microscopic dipole
traps \cite{Gaetan2009}. For our purposes, we need only consider a small
number of the internal states on which the dynamics will take place.

\begin{figure}[tb]
  \begin{center}
  \includegraphics{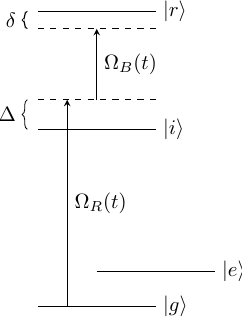}
  \end{center}
  \caption{The level scheme for a single Rydberg atom. The atom is
    driven to the Rydberg state via a two-photon transition, which
    couples the ground state \ket{g} to the excited state \ket{r}
    through the intermediate state \ket{i}. The effective Rabi
    frequencies of the transitions are $\Omega_R(t)$ for the red laser
    (which is blue-detuned by $\Delta$) and $\Omega_B$ for the blue
    laser (which is red-detuned by $\Delta + \delta$).}
  \label{fig:level_scheme}
\end{figure}

A pair of hyperfine states of the atom will encode the computational
basis, and are labelled $\ket{e} = \ket{0}$, $\ket{g} = \ket{1}$. Each
atom has its $\ket{g}$ state coupled to a highly excited Rydberg state
(which we label $\ket{r}$) via a two-photon transition through an
intermediate state $\ket{i}$. The internal level scheme with state
transitions for a single atom is shown in Fig.~\ref{fig:level_scheme}.
If two neighbouring atoms are excited to the $\ket{r}$ state, then they
interact. For the work presented we have taken a dipole-dipole potential with energy $U(r) = C_3/r^3$. However, the conclusion of the work is independent of this functional form and only the strength of the interaction at a given fixed distant of the two atoms is relevant. As a consequence our conclusions are valid also for a van-der-Waals interaction.
The interaction energy shifts the energy of the state where both atoms
are excited. When this shift is much larger than the two-photon detuning
$\delta$, the two-photon transition is far off-resonant with the
doubly-excited state, leading to a strong suppression of both atoms
becoming excited. This effect is known as the Rydberg blockade, and has
been observed experimentally \cite{Singer2004,Gaetan2009}. The effect is
shown schematically in Fig.~\ref{fig:level_scheme_2atom}.

The important point about the blockade mechanism in our case is that it
is state dependent: only if both atoms are in the ground state $\ket{g}$
will they be subject to the blockade. The potential use of this as a
mechanism for performing a quantum gate has been explored in several
papers \cite{Jaksch2000,Lukin2001,Moller2008,Brion2007}, also with Krotov pulse shape optimisation \cite{MMueller1,MMueller2}, but here we follow the
adiabatic (model B) scheme of Jaksch~\emph{et al.}, where the gate is
performed by adiabatically driving the two-atom system
\cite{Jaksch2000}.

\subsection{Gate operation in outline}

\begin{figure}[tb]
  \begin{center}
  \includegraphics{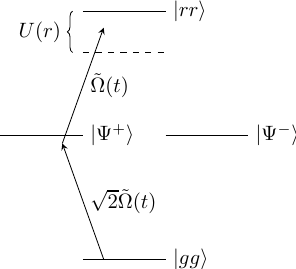}%
  \end{center}%
  \caption{The level scheme for the two-atom system (neglecting the
    internal level $\ket{i}$). The transition from the joint ground
    state $\ket{gg}$ to the super-radiant state $\ket{\Psi^+} =
    (\ket{gr} + \ket{rg})/\sqrt{2}$ is enhanced by a factor of
    $\sqrt{2}$, while there is no coupling to the sub-radiant
    $\ket{\Psi^-} = (\ket{gr} - \ket{rg})/\sqrt{2}$ state. The
    excitation of both atoms to the Rydberg state $\ket{rr}$ is
    forbidden, since the interaction energy $U(r)$ has shifted the level
    far off-resonant with the incident lasers.}%
  \label{fig:level_scheme_2atom}%
\end{figure}

There are two critical elements that allow us to synthesise the gate with
our system. The first is the blockade mechanism, which prevents
excitation to the doubly-excited $\ket{rr}$ state (where we have used
the shorthand notation $\ket{r}\otimes\ket{r} = \ket{rr}$ for the tensor
product of the state of the two atoms, which will used throughout). This
avoids unwanted mechanical effects stemming from the strong interaction
of the two Rydberg atoms, as well as reducing the time spent in the
Rydberg state, which is subject to loss. The second crucial aspect is
the super-radiant enhancement of excitation from $\ket{gg}$ as compared
with the states $\ket{ge}$ and $\ket{eg}$. In other words the Rabi frequency of this transition is enhanced by a factor $\sqrt{2}$, see Fig.~\ref{fig:level_scheme_2atom}. This results in a higher rate
of phase accumulation on the $\ket{gg}$ state during excitation in
comparison to $\ket{ge}$ and $\ket{eg}$. By carefully choosing the
excitation profile of the incident lasers, we can control these two
different accumulated phases to produce the \textsc{cz} gate.

\subsection{Hamiltonian}
The two-photon transition is driven via two lasers; one blue-detuned on
the transition from $\ket{g}$ to $\ket{i}$ by an amount $\Delta$ with a
Rabi frequency $\Omega_R(t)$, and the other red-detuned on the
transition $\ket{i}$ to $\ket{r}$ by an amount $\Delta + \delta$ with a
Rabi frequency $\Omega_B$ (see Fig.~\ref{fig:level_scheme}). In addition there is loss from the states \ket{i} and \ket{r}. The general
form of the effective Hamiltonian for our two-atom system can be written as
\begin{equation}\label{eqn:ham}
  \hat{H} = \hat{H}_r^1 + \hat{H}_r^2 + \hat{H}_{\mathit{int}}\:.
\end{equation}
The single-atom Hamiltonians are composed of both the internal and
external dynamics, such that (after the rotating wave approximation)
\begin{align}
  \hat{H}_r^j &= \hat{H}^j_S + \hat{H}^j_I + \hat{H}^j_E\:,\\ 
  \hat{H}^j_S &= (\Delta - i\gamma_i)\ketbra{i}{i} + (\delta -
  i\gamma_r)\ketbra{r}{r} \\ \label{eqn:hamint}
\hat{H}^j_I &=-\frac{\hbar \Omega_R(t)}{2}e^{i\vec{k_R}\cdot \vec{r}_j}\ketbra{g}{i} -
    \frac{\hbar \Omega_B}{2}e^{i\vec{k_B}\cdot \vec{r}_j}\ketbra{i}{r} + \mathrm{H.c}\:,\\ \label{eqn:hamext}
  \hat{H}^j_E &= (\hat{T} + V_{\mathrm{trap}})(\ketbra{g}{g} +
  \ketbra{e}{e} + \ketbra{i}{i} + \ketbra{r}{r})\:,
\end{align}
where $i = 1,2$ labels the two atoms. $\hat{H}_S$ describes the energy
splitting of the internal states along with the effective decay from those
states. This description of the loss is valid for small loss, which is the case for the optimised version of our gates. (It is a bad description in the unoptimised cases with low fidelity and high loss, but this does not affect the results in this paper.)
$\hat{H}_I$ describes the laser coupling between the internal states,
and $\hat{H}_E$ contains the kinetic and potential energy terms. The factors
$\gamma_i$, $\gamma_r$ account for an effective loss of population from the
intermediate and Rydberg levels respectively. The exponential terms in
Eq.~\eqref{eqn:hamint} are phases accumulated by the atoms as they move
through the light-field of the laser; $\vec{k_R}$ and $\vec{k_B}$ are the wavevectors of the red and blue laser
fields; $\vec{k}=\vec{k_R}+\vec{k_B}$ the wave vector of the effective two photon transition, and $\vec{r}_{j}$ is the position vector of the $j$th atom. The
interaction Hamiltonian is given simply by the dipole-dipole interaction:
$\hat{H}_{\mathit{int}} = U(r)\ketbra{rr}{rr}$.

Note that in Eq.~\eqref{eqn:hamext} we have neglected the difference in
the trapping potentials for the different internal states (in any case
we will turn off the trap when the gate is performed). Since there are
no state-dependent terms in Eq.~\eqref{eqn:hamext}, we can neglect it in
our treatment.

\subsection{Effective two-level system dynamics}
By performing an adiabatic elimination \cite{Brion2007a} of the state
$\ket{i}$, we can examine the effective three-level dynamics of the
system. This leads to the condition that $\Omega_R(t), \Omega_R, \delta
\ll \Delta$. In addition, we make a change of basis in the subspace span$\{\ket{gr},\ket{rg}\}$ such that the new basis vectors in this subspace are
\begin{align}\nonumber
  \ket{\Psi^+} \equiv \frac{1}{\sqrt{2}}\left(e^{i \vec{k}\cdot \vec{r}_1}\ket{gr} +
    e^{i \vec{k}\cdot \vec{r}_2}\ket{rg}\right)\:,\\
  \ket{\Psi^-} \equiv \frac{1}{\sqrt{2}}\left(e^{i \vec{k}\cdot \vec{r}_1}\ket{gr} -
    e^{i \vec{k}\cdot \vec{r}_2}\ket{rg}\right)\:.
\end{align}
We can now rewrite the system Hamiltonian as $\tilde{H} =
\hat{H}_R + \hat{H}_{\mathit{int}}$, where
\begin{align}
  \hat{H}_R &= \tilde{H}_r^1 + \tilde{H}_r^2 + \tilde{H}_I\:,\\
  \tilde{H}_r^j &= -\frac{\hbar \Omega_R^2(t)}{4\Delta}\ketbra{g}{g} + (\delta
  - \frac{\hbar \Omega_B^2}{4\Delta} - i\gamma_r)\ketbra{r}{r}\:,\\
  \begin{split}
    \tilde{H}_I &= -\frac{\hbar \tilde{\Omega}(t)}{2}(e^{-i \vec{k}\cdot
      \vec{r}_1}\ketbra{ge}{re} + e^{-i \vec{k}\cdot
      \vec{r}_2}\ketbra{eg}{er} + \mathrm{H.c.}) - \\ &\quad -
    \sqrt{2}\:\frac{\hbar\tilde{\Omega}(t)}{2}(\ketbra{gg}{\Psi^+} + \ketbra{\Psi^+}{gg})\label{eqn:hint}\:,
  \end{split}
\end{align}
with the effective Rabi frequency of the two-level dynamics
\begin{equation}\label{eqn:omegaeff}
  \tilde{\Omega}(t) = \Omega_B\Omega_R(t)/2\Delta\:.
\end{equation}
The magnitude of the dipole matrix elements 
\begin{gather*}
  |\bra{gg}\hat{H}_I\ket{\psi^+}| = \sqrt{2}\cdot\frac{\hbar\tilde{\Omega}}{2}\:, \quad |\bra{gg}\hat{H}_I\ket{\psi^-}| = 0\:,\\
  |\bra{ge}\hat{H}_I\ket{re}| = |\bra{eg}\hat{H}_I\ket{er}| = \frac{\hbar\tilde{\Omega}}{2}\:,
\end{gather*}
show that the state $\ket{\Psi^-}$ is not coupled to any of the other
states via the laser interaction; in other words, it is sub-radiant
\cite{Dicke1954}. In addition, the state $\ket{\Psi^+}$ is a
super-radiant state, so that the coupling between the ground state
$\ket{gg}$ and $\ket{\Psi^+}$ is enhanced by a factor of $\sqrt{2}$
compared to the transition $\ket{ge} \rightarrow \ket{re}$ ($\ket{eg}
\rightarrow \ket{er}$).

Finally, note that since $\ket{i}$ is never populated in this
approximation, so we neglect the loss term $\gamma_i$. We will also for
the moment assume that the atoms are stationary, and so the phases
accumulated from their movement in the light-field can be neglected.

\subsection{Gate operation in full}
Now we describe the operation of the gate in more detail. We start with
the initial state
\begin{equation}
  \ket{\psi(t = 0)} = \frac{1}{2}\left(\ket{gg} + \ket{ge} + \ket{eg} +
    \ket{ee}\right)\:.
\end{equation}
and define the target state at final time $T$
\begin{equation}
  \ket{\psi_G} = \frac{1}{2}\left(-\ket{gg} + \ket{ge} + \ket{eg} +
    \ket{ee}\right)\:.
\end{equation}
Note that while this seems to be a specific state transformation, as
opposed to the unitary transformation from Eq.~(\ref{eqn:cnotmat}), they
are in this case equivalent by virtue of the basis states $\{\ket{gg},
\ket{ge}, \ket{eg}, \ket{ee}\}$ not being directly coupled to one
another. Hence any initial state-dependent phases will not affect the
final outcome of the gate.

To perform the gate, the blue laser is always switched on, while the red
laser is modulated in a time-dependent fashion using an acoustic-optical
modulator. If this modulation is slow on the timescale given by
$\tilde{\Omega}(t)$ and $\delta$, then the system will adiabatically
follow the dressed states of the Hamiltonian $\tilde{H}$. Performing the
same treatment as in \cite{Jaksch2000} for our system, we similarly find
that the energy of the dressed levels adiabatically connected to
$\ket{gg}$ and $\ket{ge}$ ($\ket{eg}$) are
\begin{align}\label{eqn:ep_gg}
  \varepsilon_{gg}(t) &= \frac{1}{2}
  \left[\delta^{\prime\prime} - 4E_R(t)
    + \left(\delta^{\prime\prime2} +
      2\tilde{\Omega}^2(t)\right)^{\frac{1}{2}}\right]\:,\\ \label{eqn:ep_ge}
  \varepsilon_{ge}(t) &= \frac{1}{2}
  \left[\delta^{\prime} - 2E_R(t) +
    \left(\delta^{\prime2} +
      \tilde{\Omega}^2(t)\right)^{\frac{1}{2}}\right]\:,
\end{align}
respectively, where $\delta^{\prime} \equiv \delta - E_B + E_R(t)$ is the
effective two-photon detuning including the Stark shifts from the
adiabatic elimination of $\ket{i}$:
\begin{equation}
  E_R(t) \equiv \frac{\Omega_R^2(t)}{4 \Delta}\:, \quad 
  E_B \equiv \frac{\Omega_B^2}{4 \Delta}\:; 
\end{equation}
and 
\begin{equation}
  \delta^{\prime\prime} \equiv \delta^{\prime} -
  \frac{\tilde{\Omega}^2(t)}{2u + 4\delta^\prime-4E_R(t)}
\end{equation}
includes the additional Stark shift from the adiabatic elimination of
the $\ket{rr}$ state. The final state is
\begin{equation}
  \ket{\psi(T)} = e^{-i \phi_{gg}}\ket{gg} + e^{-i
    \phi_{ge}} \left(\ket{ge} + \ket{eg}\right) + \ket{ee}\:,
\end{equation}
where
\begin{equation}
  \phi_{gg} = \int_0^T \varepsilon_{gg}(t) \ud t\:,\quad 
  \phi_{ge} = \int_0^T \varepsilon_{ge}(t) \ud t\:,
\end{equation}
By performing state-selective qubit operations, we can realise the
\textsc{cz} gate. To see this, we first apply a state-selective phase
on the first atom: if the first atom is in the state \ket{g}, then it
receives a phase $e^{-i \varepsilon_{ge}}$. Similarly, we then apply the
same phase rotation on the second atom. After these operations, the
state becomes
\begin{equation}
  \ket{\tilde{\psi}(T)} = e^{-i (\phi_{gg} -
    2\phi_{ge})}\ket{gg} + 
  \ket{ge} + \ket{eg} + \ket{ee}\:.
\end{equation}
We can now define the gate phase
\begin{equation}\label{eqn:phase}
  \phi \equiv \phi_{gg} - 2\phi_{ge}\:.
\end{equation}
The operation of the gate is now clear: we seek to modulate
$\Omega_R(t)$ such that $\phi = (2k + 1)\pi$, $k \in \mathbb{Z}$, and
there is no remaining population in the excited states of either atom
(this is taken for granted in the adiabatic limit). The next step is to
design $\Omega_R(t)$ to achieve these conditions.


\section{Optimisation of the gate}\label{sec:opt}


\subsection{Simulation}
The system evolves in accordance with the Schr\"{o}dinger equation
($\hbar = 1$):
\begin{equation}\label{eqn:se}
  i \frac{\partial}{\partial t}\ket{\psi(t)} = \tilde{H}\ket{\psi(t)}\:.
\end{equation}
Since the Hilbert space dimension $|\tilde{H}| = N$ is relatively small,
we can simulate the gate by directly diagonalising the Hamiltonian and
using discrete time steps $dt$, such that the solution of
Eq.~\eqref{eqn:se} can be written as
\begin{equation}
  \ket{\psi(t)} = P e^{-i D dt} P^{-1} \ket{\psi(t-dt)}\:,
\end{equation}
where $P = [x_1, x_2, \dots, x_N]$ is the square matrix constructed from
the eigenvectors $\vec{x}_i$ of $\tilde{H}$, and $D =
\mathrm{diag}(\lambda_1, \lambda_2, \dots, \lambda_N)$ is a diagonal
matrix whose elements are the eigenvalues $\lambda_i$ of $\tilde{H}$.


\subsection{Optimisation method}
Many tools exist for numerical optimisation that we could employ here to
design the Rabi frequency $\Omega_R(t)$ that produces the gate.We start
by making a guess for the form of $\Omega_R(t)$, so that the evolution
is characterised by only a handful of parameters. One particular choice
is a Gaussian:
\begin{equation}\label{eqn:omega_gauss}
  \Omega_R(t) = \Omega_0 \;e^{-\left(\frac{t}{\tau}\right)^2}\:,
\end{equation}
where we will choose $\tau = 0.2T$, with total gate operation time $T$. The constant $\Omega_0$ is a
parameter which we can, in principle, choose arbitrarily (in reality
there will be constraints on this value, which we will come to
later). There are also parameters associated with the system that one
may vary, namely the atom separation $r$, the Rabi frequency $\Omega_B$,
and the detunings $\delta$ and $\Delta$. We begin by choosing a set of
reasonable values, and then we numerically optimise each of the
parameters to achieve the desired gate with a high fidelity. The
numerical method used to optimise the parameters is gradient descent.

\begin{figure}[tb]
  \begin{center}
    \includegraphics{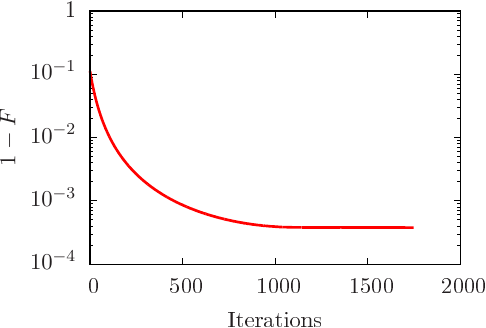}
  \end{center}
  \caption{(Color online) The decrease in infidelity of the quantum gate for the set of
    initial parameters given in Table~\ref{tab:params1}. One sees that
    the infidelity decreases monotonically until saturating in a local
    minimum of the optimisation. The final achieved infidelity was $1 - F =
    3.8\cdot 10^{-4}$.}
  \label{fig:infid}
\end{figure}

\begin{figure}[tb]
  \begin{center}
    \includegraphics[width=0.47\textwidth]{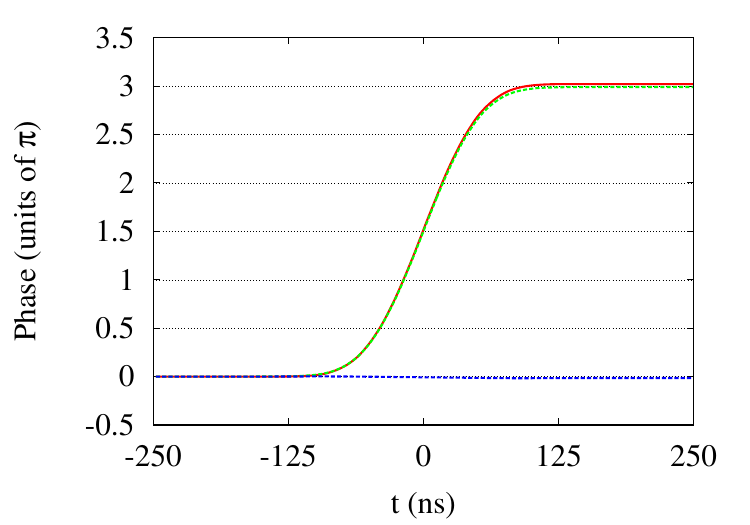}
  \end{center}
  \caption{(Color online) The dashed (green) line is the phase
    accumulation of the states $\ket{gg}$ (given by $\phi_{gg}$), the
    dotted (blue) line that of $\ket{ge}$ (given by $\phi_{ge}$), and
    the solid (red) line is the total entanglement phase $\phi$, which
    at the final time reaches $3.0$ (note that the dashed green line and 
    the solid red line overlap almost exactly). The parameters used are 
    given in
    Tab.~\ref{tab:params1}. The time is in units of ns, and so
    the duration of the gate is 500 ns.}
  \label{fig:phase}
\end{figure}

\begin{figure}[tb]
  \begin{center}
    \includegraphics{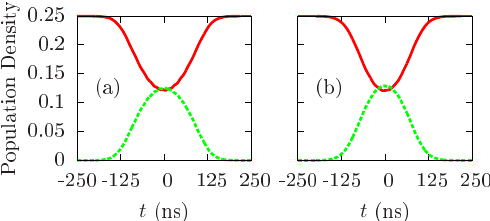}
  \end{center}
  \caption{(Color online) (a) The solid (red) line is the population of
    the state $\ket{gg}$ and the dashed (green) line is the population
    of the state $\ket{\Psi^+}$ over the duration of the gate. (b)
    Similarly to (a), the solid (red) line is the population of the
    state $\ket{ge}$ and the dashed (green) line is the population of
    the state  $\ket{re}$ over the duration of the gate. The
    populations of $\ket{eg}$ and $\ket{er}$ are respectively the same.}
  \label{fig:states_highfid}
\end{figure}

As an example, we start with the set of parameters in
Table~\ref{tab:params1}. The total time for the gate is fixed at $T=500\,$ns, and
the detuning $\delta = 0$. 
The interaction strength is chosen as $U=118\,\mathrm{GHz}$ corresponding to a separation of the atoms at a distance of 0.3 $\mathrm{\mu m}$,  when assuming dipole-dipole interaction $U(r) = 3200\mathrm{MHz}/r^3$ as in \cite{Gaetan2009} was still valid at this small distance. This is a particularly strong interaction due to the
existence of a F\"{o}rster resonance \cite{Foerster1996, Walker2005}, a
point which we will come back to later when we describe the experimental
setup in more detail. But again only the value of the interactions enters the simulations, not the distance dependency.
\begin{table}[tb]
  \begin{tabular}{@{}lcccccc@{}}
    & $\Omega_0$  & $\Omega_B$ & $\Delta$
    & $\delta$ \\ \hline
    Initial & $300$ & $300$ & $1000$ & $0$ \\
    Optimised & $304.7$ & $292.6$ & $974.8$ & 0 \\
  \end{tabular}
  \caption{A set of initial parameters that produces a gate with
    fidelity of around 89\%, and the optimised parameters that produce a
    gate with fidelity better than 99.9\%. All values are in units of 
    $2\pi\ \textrm{MHz}$}
  \label{tab:params1}
\end{table}
We define the fidelity of the gate operation as $F =
|\bracket{\psi(T)}{\psi_G}|^2$. The resultant fidelity of the gate with
these initial parameters is only around 89\%. Now we apply our
optimisation algorithm to the parameters $\Omega_0$, $\Omega_B$, and
$\Delta$. After 1600 iterations, we achieve a fidelity of better than
99.9\%, or, more precisely, and infidelity $I = (1 - F) < 3.8 \cdot 10^{-4}$. If the experimental precision of the Rabi frequencies and detuning is limited to integer MHz this infidelity slightly increases to $4.2\cdot 10^{-4}$.
The decrease in infidelity as the algorithm progresses is shown in
Fig.~\ref{fig:infid}.
Table~\ref{tab:params1} shows the final set of parameters that produced
the optimised gate. The resulting time-dependent phase accumulations
$\phi_{gg}$ and $\phi_{ge}$ are shown in Fig.~\ref{fig:phase}.
%
%
%
We see that the phase accumulated by the
$\varepsilon_{ge}$($\varepsilon_{eg}$) is exactly zero. This can be
understood by calculating Eq.~\eqref{eqn:ep_ge} for our set of
parameters: since $\delta=0$, it is straightforward to show
that $\varepsilon_{ge}(t) = 0$. The population of the relevant levels
during the gate operation are shown in Fig.~\ref{fig:states_highfid}.

It might be worrisome that the parameters in Tab.~\ref{tab:params1} do
not seem to strongly fulfil the condition $\Omega_R(t), \Omega_R, \delta \ll
\Delta$ required for the validity of the adiabatic elimination. We have,
however, confirmed that even when considering the full evolution under
$\hat{H}$ in Eq.~\eqref{eqn:ham} the population in $\ket{i}$ is
heavily suppressed, such that it may be neglected. In what follows, we
will abandon the effective model given by $\hat{H}_R$ (which provided
insight into the gate mechanism) in favour of the full treatment by
$\hat{H}$ as given by equation (\ref{eqn:ham}). This means that we include all four levels of each atom and treat the effective loss by the non-Hermitian term in equation (\ref{eqn:hamint}).


\section{Experimental considerations}\label{sec:exp}

\subsection{Level description}
We have now demonstrated the optimisation method, but we must also
consider the experimental conditions which will have an effect on the
gate fidelity. We consider the setup given in \cite{Gaetan2009}. The
gate is well suited to this system because we do not require single
atoms to be addressable, and the Rydberg blockade is relatively
strong. The reason for this is the use of a F\"{o}rster resonance that
exists in ${}^{87}$Rb \cite{Walker2005}, which comes about due to the
quasi-degeneracy of the two-atom states ($58d_{3/2}$, $58d_{3/2}$) and
($60p_{1/2}$, $56f_{5/2}$). This enhances the dipole interaction,
leading to an interaction energy $U(r) \propto 1/r^3$.

The choice of states for the levels are $\ket{g} = \ket{5s_{1/2},\ F=1,\
M_F=1}$, $\ket{e} = \ket{5s_{1/2},\ F=2,\ M_F=2}$, and $\ket{i} =
\ket{5p_{1/2},\ F=2,\ M_F = 2}$, as in reference \cite{Wilk2010}.

\subsection{Correspondence to experimental results}

As mentioned earlier, we describe the loss from both the intermediate
state $\ket{i}$ and the excited Rydberg state $\ket{r}$
phenomenologically through the decay rates $\gamma_i$ and $\gamma_r$
respectively. Examining the literature
\cite{Steck2001,Miroshnychenko2010}, we find for our setup that
$\gamma_i = 2\pi\cdot 5.75\ \mathrm{MHz}$ and $\gamma_r =
2\pi\cdot 4.8\ \mathrm{kHz}$.
\begin{figure}[tb]
  \begin{center}
    \includegraphics[width=0.47\textwidth]{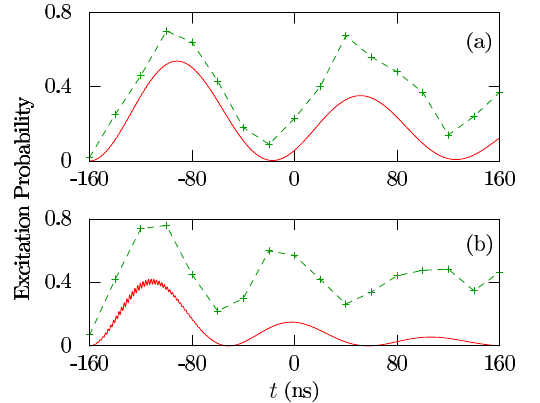}
  \end{center}
  \caption{(Color online) In (a), the solid (red) line is the simulated
    probability to excite $\ket{ge}$ to $\ket{re}$, with the
    experimental results for the same transition shown in dashed
    (green).  Similarly in (b), the solid (red) line is the simulated
    probability to excite $\ket{gg}$ to $\ket{\Psi^+}$, with the
    experimental results for the same transition shown in dashed
    (green).}
  \label{fig:comp}
\end{figure}

We have verified that the simulation corresponds to the experimental results
from \cite{Gaetan2009}. Fig.~\ref{fig:comp} shows the agreement between the
simulated Rabi oscillations of the $\frac{1}{2}\left(\ket{er} +
\ket{re}\right)$ state and the $\ket{\Psi^+}$ state. While the fit is not
exact, we do reproduce the correct frequency of oscillation, as well as an
indication of the typical decay from the intermediate level. We also find a
relative difference in frequency of the two oscillations of a factor $\sim
\sqrt{2}$, as expected from the theory. The discrepancies between theory and
experiment arise from experimental imperfections which are not taken into
account in our model, namely laser fluctuations in both power and frequency of
the lasers, which lead to some dephasing. Our phenomenological loss model also
does not account for the possibility that an atom can decay to $\ket{g}$ from
where it may be repumped, which partially accounts for the discrepancy in
total population.

\subsection{Typical experimental parameters}

While it would be ideal if the gate parameters described in the last section
could be immediately applied in the experiment, the reality is that there are
certain experimental limitations that prevent us from doing so. Firstly, the
power of both lasers have certain maximum values: $\Omega_R$ has a maximum
operating value of $1\ \mathrm{GHz}$, while $\Omega_B$ is limited to $120\
\mathrm{MHz}$. Secondly, we do not have the freedom to modulate the laser
power as we like; acoustic-optical modulators (AOMs) control the power of the
laser beams incident on the atoms, and they have limits on the rate at which
the intensity of the beam can be changed. (In any case, since the gate is
adiabatic, we expect that the final result will not depend very strongly on
the exact shape of the excitation as long as the area of the pulse is
preserved.) The `rise-time' (the time it takes to increase the laser power
from zero to its maximum) is typically in the range $200$--$400\,\mathrm{ns}$.
We take the profile of this rise-time to be Gaussian, in agreement with the measured pulse shape on the experiment. Lastly, due to the
separate dipole traps, the minimum distance between the two Rydberg atoms is
limited to $3\,\mathrm{\mu m}$ or above. A final condition is put on the
detuning $\Delta$ of the red laser: this should be less than $500\,\mathrm{MHz}$ due to experimental constraints (although this is not a
stringent condition).

There is also an additional probability of loss when an atom is
excited to the Rydberg level: when excited, the motional wavepacket
starts to spread, so that when the dipole trap is reapplied, there is a
finite probability that the atom is lost (this is actually used as a
method of detection in the experiment). This motivates us to limit the
time spent in the Rydberg state.

\subsection{Optimising the gate for experiment}

With these limitations, we now see that the gate parameters from
Tab.~\ref{tab:params1} are not feasible in our chosen experimental
setup. We must now start with a new set of parameters and run the
optimisation again to see to what extent the gate is still
implementable. Given the discussion above, we are motivated to make the
following changes to our parameters.

\begin{itemize}
\item The gate should be performed as quickly as possible, meaning that
  the effective Rabi frequency $\tilde{\Omega} \propto \Omega_R$,
  $\Omega_B$ should be made large. This implies that we should choose
  $\Omega_R$ and $\Omega_B$ close to their maximum values. (This has the
  additional advantage that we spend less time in excited Rydberg
  levels, improving the probability of recapture.) Since the constraint is stricter for $\Omega_B$ and we want to fully exploit the experimentally feasible maximum power we set $\Omega_B$ to $120\,$MHz and instead optimise the gate operation time $T$.

\item To avoid excitation of the lossy $\ket{i}$ state, we need
  to keep the red laser far-detuned, ideally around $500\
  \mathrm{MHz}$. However, as can be seen in Eq.~\eqref{eqn:omegaeff},
  increasing the detuning will reduce the effective Rabi frequency,
  which makes achieving the gate in a short time more difficult.

\item To make maximum use of the laser power, we change the
  shape of the pulse from a simple Gaussian to a 'flat-top', given by
  \begin{equation}\label{eqn:omega_flattop}
    \Omega_R = \begin{cases}
      \Omega_0 \exp\left[-\frac{(t -\tau )^2}{\tau^2/8}\right] & t \leq \tau\:, \\
      \Omega_0 \exp\left[-\frac{(t -(T-\tau) )^2}{\tau^2/8}\right] & t \geq T-\tau\:, \\
      \Omega_0 & \mathrm{otherwise}.
    \end{cases}
  \end{equation}
  Here, the Rabi frequency increases with a Gaussian profile to the
  maximum $\Omega_0$ in a time $\tau$ (the rise-time), followed by a
  period of constant Rabi frequency for a time $T - 2\tau$, and then
  finally a reduction along a Gaussian profile to zero, again in a time
  $\tau$. This gives us the freedom to have the laser at full power for the longest
  time possible, which in turn causes us to accumulate the
  time-dependent phase more rapidly. It turns out, however, that the best choice for $\tau$ is to extend the rise-time to $\tau=T/2$, thus choosing a Gaussian profile.

\item Since the minimum distance between the atoms in the experiment is
  $r_{\mathrm{min}} = 3.0\ \mathrm{\mu m}$, we will use this value in
  what follows to ensure we are as deep in the blockade regime as
  possible.

\end{itemize}

\begin{figure}[tb]
  \begin{center}
    \includegraphics[width=0.47\textwidth]{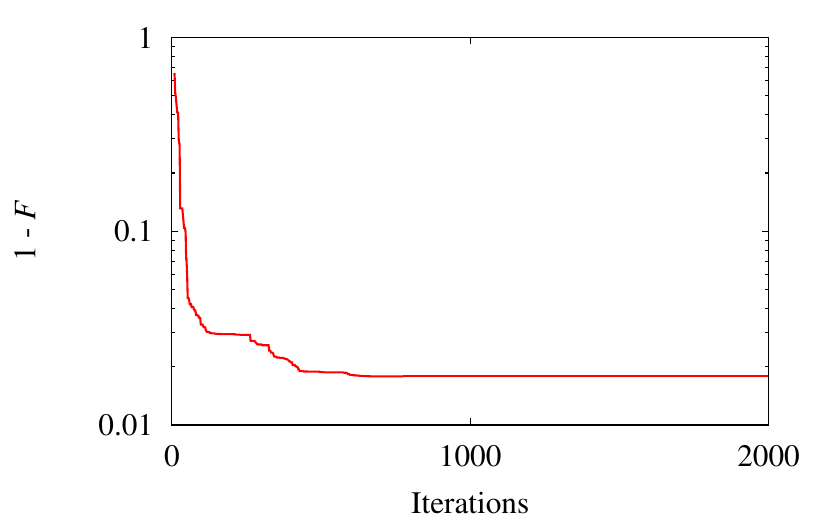}
  \end{center}
  \caption{The decrease in infidelity of the quantum gate for the set of
    initial parameters given in Table~\ref{tab:params2}. The stepwise decrease of the current minimum illustrates how restarting the simplex can result in finding another (deeper) local minimum until it reaches the (supposed) global minimum. The final achieved infidelity was $1 - F =
    1.8\cdot 10^{-2}$.}
  \label{fig:infid2}
\end{figure}

\begin{figure}[tb]
  \begin{center}
    \includegraphics[width=0.47\textwidth]{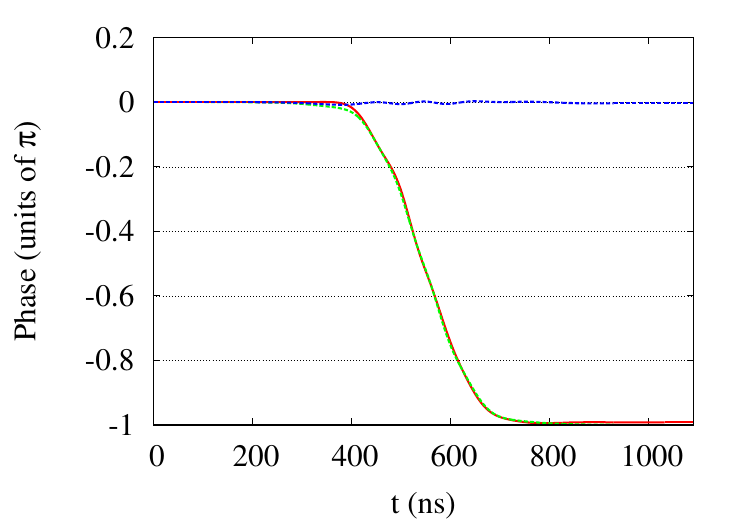}
  \end{center}
  \caption{(Color online) The dashed (green) line is the phase
    accumulation of the states $\ket{gg}$ (given by $\phi_{gg}$), the
    dotted (blue) line that of $\ket{ge}$ (given by $\phi_{ge}$), and
    the solid (red) line is the total entanglement phase $\phi$, which
    at the final time reaches $-1.0$. The parameters used are given in
    Tab.~\ref{tab:params2}. The time is in units of ns, and so the
    duration of the gate is $1.0922\,\mu$s.}
  \label{fig:phase2}
\end{figure}

\begin{figure}[tb]
  \begin{center}
    \includegraphics[width=0.47\textwidth]{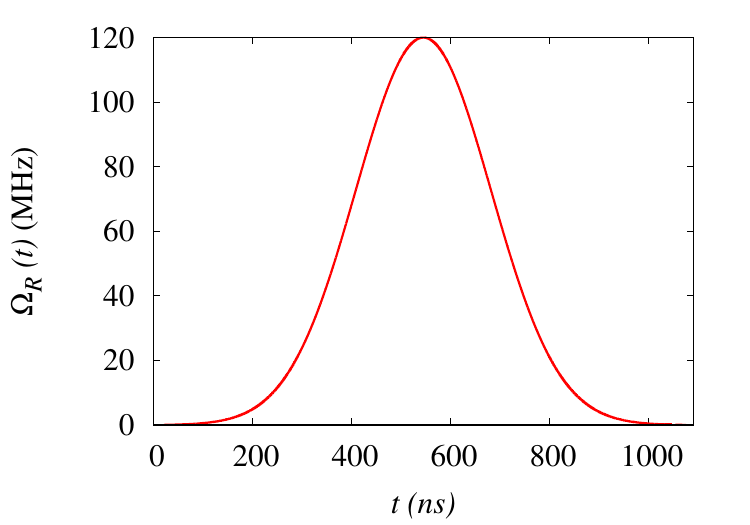}
  \end{center}
  \caption{(Color online) The pulse shape for the Rabi frequency
    $\Omega_R(t)$ for the parameters in Tab.~\ref{tab:params2}. The
    shape is a `degenerated flat-top', as given in Eq.~\eqref{eqn:omega_flattop}, i.e. a Gaussian profile.}
  \label{fig:pulse2}
\end{figure}

\begin{figure}[tb]
  \begin{center}
    \includegraphics[width=0.47\textwidth]{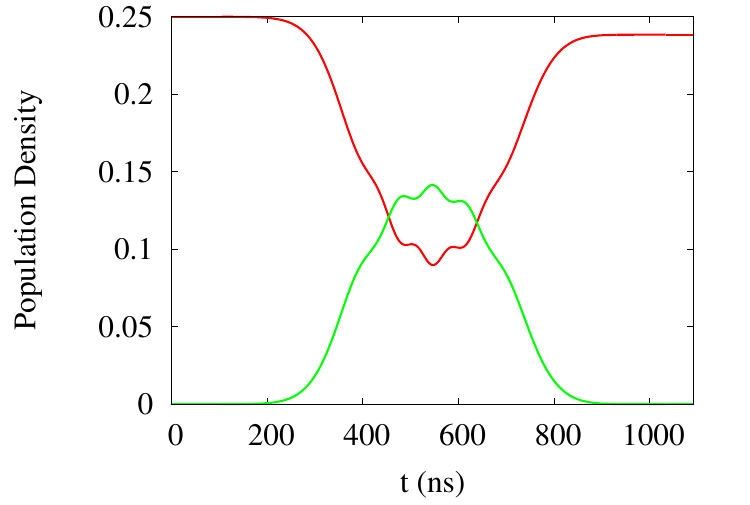}
  \end{center}
  \caption{(Color online) The solid (red) line is the population of the
    state $\ket{gg}$ and the dashed (green) line is the population of
    the super-radiant state $\ket{\Psi^+}$ over the duration of the gate.}
  \label{fig:ggpop}
\end{figure}

\begin{figure}[tb]
  \begin{center}
    \includegraphics[width=0.47\textwidth]{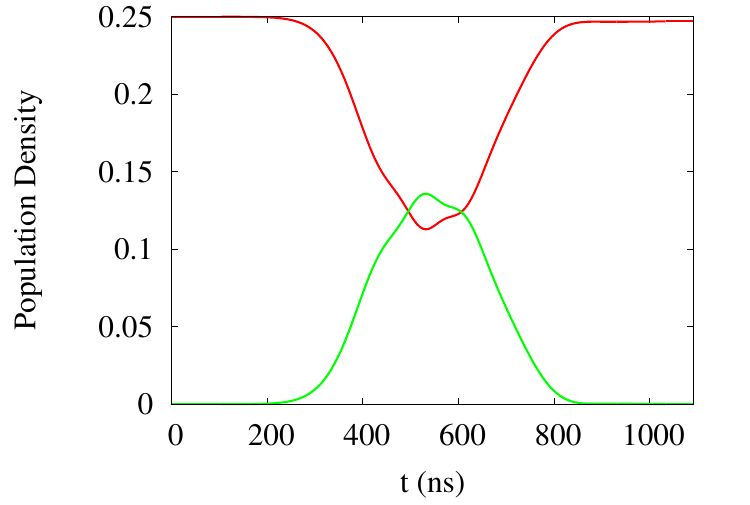}
  \end{center}
  \caption{(Color online) The solid (red) line is the population of the
    state $\ket{ge}$ and the dashed (green) line is the population of
    the state $\ket{re}$ over the duration of the gate. The populations
    of $\ket{eg}$ and $\ket{er}$ are respectively the same.}
  \label{fig:egpop}
\end{figure}

Based on these considerations, we try the set of parameters given in
Tab.~\ref{tab:params2}.
\begin{table}[tb]
  \begin{tabular}{@{}lccccccc@{}}
    & $\Omega_0$ & $T$ & $\Delta$
    & $\delta$ \\ \hline
    Initial & $50$&$1$&$500$&$-1$\\
    Optimised & $129.6$ & $1.1$ & $-703.7$ & $0.1$\\
  \end{tabular}
  \caption{A set of initial parameters, within experimental constraints, 
    that produces a gate with
    fidelity of around 52\%, and the optimised parameters that produce a
    gate with fidelity better than 98\%. Rabi frequencies are in units of $2\pi$ MHz, time in units of $1\,\mathrm{\mu s}$.}
  \label{tab:params2}
\end{table}
This time we use the Nelder Mead simplex algorithm to optimise the gate parameters since the physics here make it more challenging to perform a high fidelity gate and we want to avoid local minima. As before, we can examine the convergence of the
optimisation (Fig.~\ref{fig:infid2}), the accumulation of the
entanglement phase (Fig.~\ref{fig:phase2}), and the shape of the Rabi
frequency $\Omega_R(t)$ (Fig.~\ref{fig:pulse2}). 
The final gate fidelity is only around 98\%, mainly due to losses, resulting in a final population norm of 0.983. This loss mainly occurs in the time-evolved $\ket{gg}$ state as can be seen in Fig.~\ref{fig:ggpop}, while the results in Fig.~\ref{fig:egpop} show that there is less loss when time-evolving $\ket{eg}$, since here only one atom is excited. Reducing the precision of Rabi frequencies and detunings to integer MHz increases the infidelity from $1.8\cdot 10^{-2}$ to $2.2\cdot 10^{-2}$

We believe that this result is close to the optimal case for this system
and gate implementation given the experimental limitations. Only if we release the constraint on $\Omega_B$ we can further improve the fidelity. We have checked that we can cross the 99\,\% fidelity threshold at about $\Omega_B=200\,$MHz which at the moment needs still improvement of experimental equipment.

\subsection{Movement of the atoms in the light-field}

Until now, we assumed that the atoms were stationary in the dipole
traps. In reality, the atoms are Doppler-cooled to around $75\
\mathrm{\mu K}$. During the laser excitation, the trapping fields are
switched off, allowing the atoms to move freely in any direction in the
plane perpendicular to the trapping field. The terms $\vartheta_i \equiv
\arccos(\vec{k}\cdot\vec{r}_i) \in [0,\pi]$ from Eq.~(\ref{eqn:hint}) then
produce additional independent phases on each atom. Since we don't know
{\em a priori} in which direction the atoms will move with respect to the
light field, the phase difference between the two atoms is essentially
random.

\begin{figure}[tb]
  \begin{center}
    \includegraphics[width=0.47\textwidth]{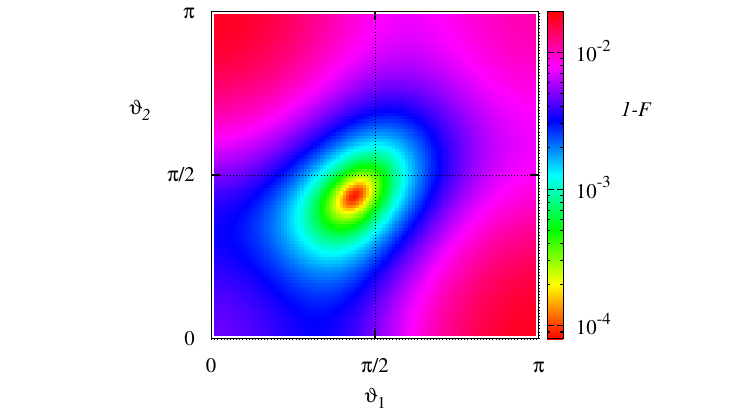}
  \end{center}
  \caption{(Color online) The effect of the motional phases of the atoms
    on the final gate fidelity for the parameters in
    Tab.~\ref{tab:params1}. Here, the effect of the motional phases is
    not too large, since we perform the gate over a very short time and
    we do not significantly excite the atoms.}
  \label{fig:lfhf}
\end{figure}

\begin{figure}[tb]
  \begin{center}
    \includegraphics[width=0.47\textwidth]{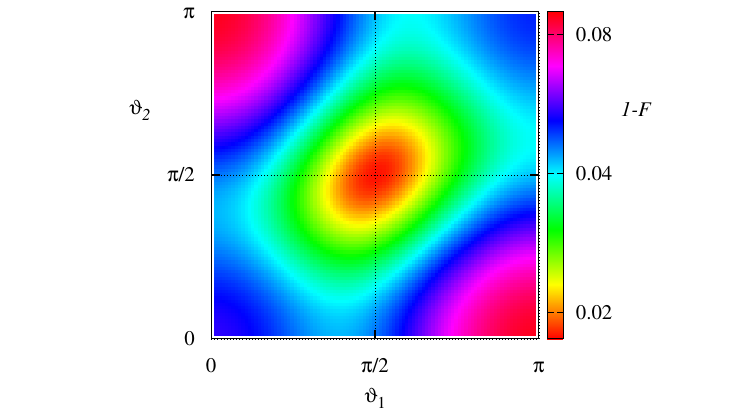}
  \end{center}
  \caption{(Color online) The effect of the motional phases of the atoms
    on the final gate fidelity for the parameters in
    Tab.~\ref{tab:params2}. Here, the effect of the motional phases is
    detrimental to the gate fidelity.}
  \label{fig:lfll}
\end{figure}

Figures~\ref{fig:lfhf} and \ref{fig:lfll} show the effect of this phase
on the final fidelity of the gate for the sets of
optimal parameters in Tabs.~\ref{tab:params1} and \ref{tab:params2}
respectively. In the first implementation of the gate, we see that the
effect of the motional phase on the gate fidelity is, at worst, a drop
in infidelity from $10^{-4}$ to $10^{-2}$. In the second case, the
effect is about from $10^{-2}$ to $10^{-1}$.

\begin{figure}[tbp]
  \begin{center}
    \includegraphics[width=0.47\textwidth]{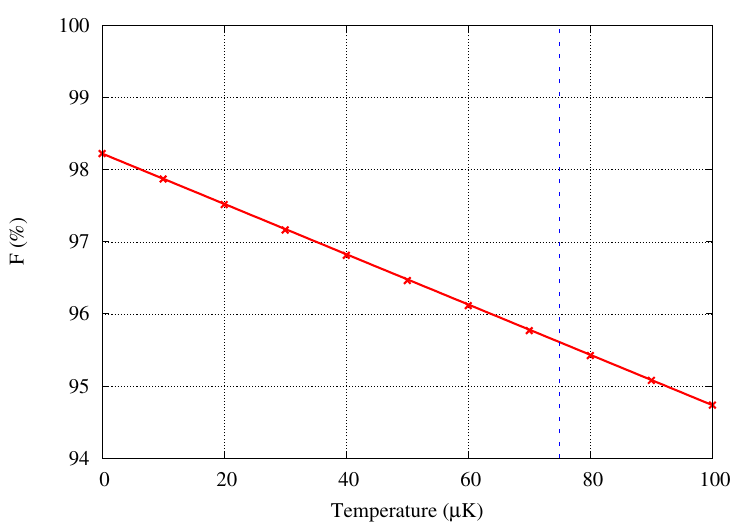}
  \end{center}
  \caption{(Color online) The final fidelity of the gate plotted against
    the average temperature of the atoms. Each (green) point is an
    average over 10,000 realisations of $\vartheta_1$ and $\vartheta_2$
    chosen randomly in the range $[0,\pi]$. The (red) solid line is a
    linear fit to the points. The (blue) dashed line shows the current
    temperature of 75 $\mu$K.}
  \label{fig:temp_vs_fid}
\end{figure}

To actually perform the gate in practice under these conditions, we
would have to cool the atoms much closer to the motional ground
state. This would reduce the distance that the atoms move in the
light-field, and hence the amount of phase that they collect. The
dependence of the fidelity on the average temperature of the atoms is
given in Fig~\ref{fig:temp_vs_fid}.
Thus by reducing the temperature by a factor of around five will increase the
fidelity to around 97\%. This is experimentally realistic as demonstrated in 
reference \cite{Tuchendler2008}.


\section{Conclusion}\label{sec:conc}

We have investigated the implementation of the Rydberg two-qubit entangling
gate from Jaksch \emph{et al.}\cite{Jaksch2000} in the experiment outlined in
\cite{Gaetan2009}. We applied a direct search (gradient descent and Nelder Mead simplex) control algorithm to
find optimal sets of experimental parameters that produced a gate with low
loss from the intermediate level while still achieving a fidelity of around
98\%. The main source of error was found to be the random phase accumulated by
the atoms in the light field.

While the system seems ideally suited for this scheme, the experimental
constraints still limit the fidelity of the gate. The most notable source of error comes from the movement of the
atoms in the light field, which could be minimised in future experiments
by cooling the atoms further. Eliminating this source of error should
allow gate fidelities of around 99\% which, while not quite good enough
to allow quantum computation (even with error correction), would be a
significant step forward for the realisation of quantum computation with
neutral atom systems.

It is worth pointing out that while we investigated the gate scheme of
\cite{Jaksch2000}, there are some alternative schemes that could be
implemented in our setup, most notably perhaps the scheme of
\cite{Moller2008} which uses a \textsc{stirap} pulse sequence to excite
the atoms and also the individual addressing of atoms as analysed in detail in \cite{Zhang2012}. We have also not investigated allowing the Rabi frequency of
the blue laser $\Omega_B$ to modulate in time which could lead to a more
robust gate implementation.

We acknowledge financial support by the EU under the contracts
MRTN-CT-2006-035369 (EMALI), IP-AQUTE, MALICIA, and SIQS; from the German SFB TRR21 and QuOReP. We
also acknowledge computational resource provided by the BWgrid.

%

\end{document}